\documentclass[showpacs,showkeywords]{elsart}
\usepackage{graphicx}
\usepackage{natbib}
\usepackage{latexsym}
\usepackage{amssymb}

\begin{document}
\begin{frontmatter} 



\title{Diagrammatic Computation of the Random Flight Motion}
\author{S. T. Hatamian}
\address{Mathematicus Laboratories, 25 New York Ave, Box 1296,
Sound Beach, NY 11789-0972 }
\date{\today }
\ead{tsh@MathematicusLabs.com}  

\begin{abstract}
We present a perturbation theory by extending a prescription due
to Feynman for computing the probability density function for the
random flight motion. The method can be applied to a wide variety
of otherwise difficult circumstances. The series for the exact
moments, if not the distribution itself, for many important cases
can be summed for arbitrary times. As expected, the behavior at
early time regime, for the sample processes considered, deviate
significantly from diffusion theory; a fact with important
consequences in various applications such as financial physics. A
half dozen sample problems are solved starting with one posed by
Feynman who originally solved it only to first order. Another
illustrative case is found to be a physically more plausible
substitute for both the Uhlenbeck-Ornstein and the Wiener
processes. The remaining cases we've solved are useful for
applications in regime-switching and isotropic scattering of light
in turbid media. It is demonstrated that under isotropic random
flight with invariant-speed, the description of motion in higher
dimensions is recursively related to either the one or two
dimensional movement. Also for this motion, we show that the
solution heretofore presumed correct in one dimension, applies
only to the case we've named the "Shooting Gallery" problem; a
special case of the full problem. A new class of functions dubbed
``Damped-exponential-integrals" are also identified.
\end{abstract}

\end{frontmatter} 
\maketitle
\section{introduction}

The understanding of Brownian-motion has been of fundamental
import on many levels in pure and applied physics. For example,
the propagation of light in gaseous media has long been important
in atmospheric and stellar physics\cite{Chandra}. In
chemical-physics the topology of polymers\cite{kleinert} is of
great interest. Yet, unlike topics of comparable ubiquity,
Brownian-motion has immediate applications in the human realm,
ranging from the physics of finance\cite{Osborne,PhysicaA_v269},
to medical imaging\cite{OptTomo}, to name only a very few.

The diversity of the applicants has resulted in a plethora of
methodologies delineating the evolution of this ubiquitous topic.
Among these techniques, the path integral formulation of the
stochastic movement, while often more imaginative, is fairly
under-represented. Meanwhile, the standard methodologies based on
differential equations, often lead to difficult and unintuitive
equations as soon as one gets away from the limiting situations
such as infinitesimal mean-free-paths, or long-time regimes.

For example, the propagation of light through a stochastic medium
is traditionally described in the context of astrophysics by a
Boltzmann transport equation for the distribution of intensity
$I(\textbf{r},\Omega ,t)$ in a heuristic Radiative transfer theory
\cite{Chandra2}. However,since the general analytic solutions are
unknown, one resorts to the diffusion approximation which can be
shown to arise out of the Radiative transport equation in the
limit of large length scales $X \!>>\! L_0$, where $L_0$ is the
transport mean-free-path of light in the
medium\cite{Chandra2,Ishimaru} .

In this same vein, there is considerable interest in the
description of multiple light scattering at small length scales
($X \!\sim\! L_0$) and small time scales $\lambda T\!\sim\! 1$
where $1/\lambda $ is the transport mean free time, both from the
point of fundamental physics \cite{Kop}, and from the point of
many applications such as medical imaging, etc. \cite{Colak,Wang}.
It has been experimentally shown that the diffusion approximation
fails to describe phenomena at time scales where $\lambda T\!
\lesssim\! 10$ \cite{Yoo}. Moreover the diffusion approximation,
which is strictly a Wiener process, for the spatial coordinates of
a particle is physically unrealistic. It holds only in the limit
of the mean free path $L_0\!\rightarrow\! 0$ and/or
$\lambda\rightarrow \infty$ or the speed of propagation $c\!
\rightarrow \!\infty$, while keeping the diffusion coefficient ($D
\sim c^2/\lambda$) constant.

It is somewhat surprising that developing better alternatives to
the diffusion approximation has had to wait until the 1990's. For
a particle moving with invariant speed in a one-dimensional
disordered medium, it has been known for decades that the
probability distribution function $P(X,T)$ for the displacement
satisfies the telegrapher equation exactly\cite{Goldstein}.
However, generalizations of the telegrapher equation to higher
dimensions \cite{Durian} have been shown not to yield better
results than the diffusion approximation \cite{Porra,Paasschens}.
Indeed we shall show (in section VI) that even the forementioned
one-dimensional exact solution does not fully describe isotropic
scattering, and additional diagrams are necessary to allow true
isotropy. For a concise account of the recent work in light
scattering work see \cite{Hindus}.

In a similar mooring is the description of price movements in
openly traded markets, where the diffusion approximation has been
taken as gospel since the inception of theoretical finance.
Meanwhile it has been known for over a century\cite{Pareto} that
the actual probability distributions of the market prices deviate
significantly from those expected based on the diffusive (Wiener)
processes. More directly, there has never been any concrete
empirical reasons to believe a property such as
$\lambda\!\rightarrow\!\infty$ holds for the movement of openly
traded market prices. With $\lambda\!\sim\! 10^{21}/sec$ for the
typical gasses which concerned the developers of the early physics
literature\cite{Chandra}, the approximation was more than
adequate. On the surface it appears that this notion was taken
from physics without proper care to ask if the underlying
assumptions are satisfied in the movement of market prices. We
submit that addressing this ``misapplication" is well overdue. The
recent wave of exploratory work on the notion of ``regime
switching" \cite{MarkovSrch,Veronesi} displays the burgeoning
dissatisfaction with the gospel of diffusion in econophysics in
general.

Elegant prescriptions for the solution of the Fokker-Planck
equation exist with \cite{Kleinert3} and without \cite{Risken,
Martin} resort to path-integral techniques which address the
problem of classical stochastic probability density functions for
arbitrary parameters and time scales. However, techniques
generally become cumbersome when applied to motion not prescribed
by analytic equations. Alas, many problems, particularly outside
the realm of physical systems, involve discrete and/or non-smooth
motion which are easy enough to conceive but not easily expressed
analytically. The method we outline is arguably better suited for
this type of problem than other, more sophisticated, theories.

In this report we present a technique based on a method invented
by Feynman\cite{FeynPIBook}, allowing notions of Feynman diagrams
for the random-flight movement. The technique of path
integration\cite{Wiener}, also known as sum over histories, is
particularly attractive in the case of discrete stochastic motion
because of the vivid imagery it evokes, literally illustrating the
mathematical process\cite{papadop}.

In section-\ref{s-FeynPrescript} we summarize the groundwork on
which we will build our method. In section-\ref{s-GSProcess} we
present the full solution to an illustrative problem set up by
Feynman\cite{FeynPIBook} but only solved for small displacements.
Section-\ref{s-Noise} illustrates how extraneous features such as
noise or force fields can easily be integrated in the
``propagator". As the final practice problem,
section-\ref{s-GRProcess} presents the solution of an important
variation on the problem of section-\ref{s-GSProcess}. This
variant can be considered a microscopically plausible alternative
to the UhlenBeck-Ornstein process. The ensuing two sections,
consider the problem of random flights under invariant speed,
where the majority of the results presented are novel. We then
conclude with two specific applications of selected results in
modelling flexible (polymer) chains and stock prices.

\section{Feynman's prescription for stochastic processes}\label{s-FeynPrescript}
In this section we review the prescription\cite{FeynPIBook} for
computing the probability density function for a random process
using functional integrals. Suppose that $f(t)$ describes a
realization of a random process as function of time. Called a
``probability functional", $P[f(t)]$, then gives the probability
density of a given realization. If we think of $f(t)$ as an
ordered collection of discrete values $f_1,f_2,...,f_n$ in the
limit of $n\!\!\rightarrow\!\!\infty$, the probability of finding
a realization $f(t)$ is given by $P[f(t)]\mathcal{D}f(t)\equiv
\lim_{n\rightarrow\infty} P(f_1)P(f_2)...P(f_n)df_1df_2..df_n$
within a hyper-neighborhood given by the latter expression. The
conjugate ``characteristic functional" is computed by:
\begin{equation}\label{PhiDefn}
\Phi[k(t)]\equiv \frac{ \int{ e^{i\!\int{k(t)f(t)dt}}}
P[f(t)]\mathcal{D} f(t)}{\int P[f(t)]\mathcal{D}f(t)}.
\end{equation}
Feynman\cite{FeynPIBook} showed that despite the challenging
appearance of this expression, it might be possible that under
certain practical conditions the discretized form of
$P[f(t)]\mathcal{D}f(t)$ can yield to either term by term
integration, or known forms of path-integrals. Suppose that each
random event has a time signature given by $g(t)$. Then the full
realization of $n$-events over a fixed time interval $T$ is simply
$f(t)=\sum_{j\,=1}^n g(t-t_j)$. If the events are random and
uniformly distributed in time, then the probability of each event
occurring within the interval $dt$ is $dt/T$. Inserting $f(t)$ and
$P[f(t)]$ into Eq.(\ref{PhiDefn}) and noting the identical form of
each integral we get:

\begin{eqnarray}\label{e-Phi_ng}
\lefteqn{\nonumber \Phi_{ng}[k(t)]= \int_0^T \exp{\left[ i
\sum_{j\,=1}^n \int k(t)g(t-t_j)dt \right]} \;\;\frac{dt_1}{T}
\ldots \frac{dt_n}{T}} \\ &&
\;\;\;\;\;\;\;\;\;\;\;\;\;=\left(\int_0^T e^{i\!\int
k(t)g(t-\tau)dt}\frac{d\tau}{T}\right)^n.
\end{eqnarray}
where we have used $\tau$ as a stand in for any given $t_j$. For
typical applications, the shape or at-least the amplitude of each
event is often not fixed. We therefore consider the case
$g(t)\equiv au(t)$ where the shape $u(t)$ of the event is fixed
but its amplitude $a$ varies according to some probability density
$p(a)$. We must now average $\Phi_{ng}[k(t)]$ over all values of
$a$. Being that the latter is of form $(\phi_g[k(t)])^n$, and that
the events are independently distributed, we can carry out this
average over all \emph{g}'s separately for each component
$\phi_g$:
\begin{equation}
\phi\equiv <\!\phi_g\!>_g=\!\int_0^T \!\frac{d\tau}{T}\!\! \int
\exp{\left[i a\!\! \int\! k(t)u(t-\tau)dt\right]}p(a)da.
\end{equation}
Observe that the inner expression is a fourier transform of $p(a)$
where the frequency variable is the integral in the exponent. The
fourier-transform being a characteristic-function, will be
designated as $W[\int k(t)u(t-\tau)dt]$. Finally, if the number of
events over $T$ is governed by a Poisson distribution, we can
readily average over all possibilities of $n$:
\begin{equation}
\Phi[k(t)]=\sum_n \phi^n \frac{\bar{n}^n}{n!} e^{-\bar{n}}.
\end{equation}
Here $\bar{n}=\lambda T$ is the average number of events expected
over $T$. The sum is simply an exponential:
\begin{equation}\label{FBEqn1}
\Phi[k(t)]=e^{-(1-\phi)\bar{n}}=}  \exp{
\left[-\lambda\int_0^T\left( 1- W\!\!\left[\int
k(t)u(t-\tau)dt\right]\right)d\tau\right].
\end{equation}
An important special case of the process described by
Eq.(\ref{FBEqn1}) is when the signal-event has an extremely short
duration in time: $u(t)=\delta(t)$. Then the characteristic
functional becomes:
\begin{equation}\label{FBEqn2}
\Phi[k(t)]=\exp{ \left[-\lambda\int_0^T(1-
W[k(\tau)])\,d\tau\right]}.
\end{equation}

\section{The Gaussian Scattering Process}\label{s-GSProcess}
Using the characteristic Eq.(\ref{FBEqn2}) we can compute various
moments of the signal profile $f(t)$. However, the true utility of
this approach is to discover the resulting distributions of
dependent variables which are \emph{driven} by the events in
$f(t)$. In this, and the following two sections,  we will
investigate cases of movement of free particles of unit mass,
which undergo a prescribed change in their velocity for every
event in $f(t)$. Initially the particles have a delta-function
distribution in both speed and spatial spread. In between events,
the movement of the particle is governed by its appropriate
equation of motion. Our objective is to find the resulting spatial
distribution of particles after a given time $T$.

At every event the particle acquires a change in momentum which is
a random selection from a gaussian distribution of spread $s$ and
zero mean. The characteristic function $W(\omega)$ of the gaussian
random force-function profile is simply another gaussian. When
inserted into Eq.(\ref{FBEqn2}) we have:
\begin{equation}
\Phi[k(t)]=e^{-\lambda T}\exp{ \left[\lambda\int_0^T
e^{-k(t)^2s^2/2}dt\right]}.
\end{equation}

As discussed in the previous section, the probability density of
the force function $P[f(t)]$ is given by the fourier transform of
$\Phi[k(t)]$, as given by Eq(\ref{PhiDefn}):
\begin{equation}
\!\!\!\!\!\!\!\!\!\!\!\!P[f(t)]\!=\!\int\!\!\Phi[k(t)] e^{-i\int
k(u)f(u)du} \mathcal{D} k= e^{-\lambda T} \!\!\! \int
\!\!e^{\lambda\int_0^T e^{\!-k(t)^2s^2/2} dt} e^{-i\int_0^T
\!k(u)f(u)du}\mathcal{D} k
\end{equation}

In order to proceed, we discretize the time interval $[0,T]$ into
$N$, regular small intervals of duration $\eta$. With the
understanding that $N\!\!\rightarrow\!\infty$ we can write:
\begin{equation}
P[f_i]=e^{-\lambda T}\prod_i^N \int \frac{dk_i}{2\pi}\exp{
\left[\lambda \eta e^{{-k_i}^2s^2/2}\right]}e^{-i\eta k_i f_i}.
\end{equation}
Finally, we expand the first exponential in a Taylor series and
carry out the individual fourier transforms:
\begin{equation}\label{EqnPfi}
P[f_i]=e^{-\lambda T}\prod_i^N \left[ \delta(f_i\eta) +
\frac{\lambda \eta}{s\sqrt{2\pi}} e^{-{(f_i \eta)}^2/2s^2} +
\ldots \right]
\end{equation}
whereupon we recognize $f_i \eta\equiv I_i$ as the impulse
imparted at time $t_i$. The leading (zeroth-)order in $\eta$ is a
product of $\delta$-functions which we'll call a
$\delta$-functional $\delta[I(t)]$. The value of the functional is
zero unless the function $I(t)$ is zero for all $t$. The
zeroth-order term
\begin{equation}
P_0[I(t)]=\delta[I(t)]
\end{equation}
corresponds to the ballistic path where the particle experiences
no collisions and hence experiences no impulses. The $O(\eta)$
terms are comprised of all $\delta$'s, except at time $t_i$ where
a single term of $O(\eta)$ contributes. There are $N$ such terms
(one for each $t_i$) comprising a Riemann sum. In the limit of
large $N$ this sum becomes the integral:
\begin{equation}
P_1[I(t)]=\frac{1}{s\sqrt{2\pi}}\int_0^T dt\; e^{-{I(t)}^2/2s^2}
\bar{\delta}[I(t)] .\end{equation} The functional
$\bar{\delta}[I(t)]$ is defined as the
($\lim_{N\rightarrow\infty}$) product of $\delta (I(t_i))$ for all
$t_i$ except $t_i=t$. This term describes the path where there is
only a single collision at time $t$. It is easy to show that if
the scattering profile had an offset, then it would change the
exponent of the integrand to $I(t)-I_0$. Higher order terms in
$\eta$ are easily obtained:
\begin{equation}\label{EqnPn}
P_n[I(t)]=\prod_{i=1}^n\frac{1}{{s\sqrt{2\pi}}}\int_0^{t_{i+\!1}}
dt_i e^{-{I(t_i)}^2/2s^2} \bar{\delta}[I(t_1),...,I(t_n)]
\end{equation}
where $t_{n+1}\equiv T$. Upon collecting the orders we get the
expression for the probability density functional of
Eq(\ref{EqnPfi}):
\begin{equation}\label{Eqn3-3}
P[I(t)]=e^{-\lambda T} \sum_{n=0}^\infty \lambda^n P_n[I(t)].
\end{equation}
In order to obtain the probability density of a given output
position $P(X)$ we must first derive $P[x(t)]$ from $P[I(t)]$. We
then will sum over all paths $x(t)$ which satisfy the boundary
conditions: $x(t=0)=x_0$ ; $x(t=T)\equiv X$; $\dot{x}(t=0)= v_0;
\dot{x}(t=T)=V$.

The final sum over paths can easily be done because of the insight
we have acquired by recognizing the various orders in the series
of Eq.(\ref{Eqn3-3}) as paths with a given number of events. The
zeroth order term is the easiest; it is simply the sum over all
paths which experience no impulses, of which there is only one.
However, to proceed more formally we will utilize the following
observation\cite{FeynPIBook}. As long as the relation between $I$
and $x$ is linear (e.g. $I\propto\ddot{x}$), we can be sure that
any Jacobian resulting from the change of variable:
\begin{equation}\label{FeynObs}
P_n[I(t)]\mathcal{D}I=P_n[x(t)]\mathcal{D}x
\end{equation}
is a constant, and if skipped, affects \emph{only} the
normalization of the final answer. The normalization of the
resultant distribution must be ensured regardless. Thus, from
Eq.(\ref{EqnPn}), and (\ref{FeynObs}):
\begin{equation}\label{e-GSpropagator}
P_0(X,T)=\int \delta[I(t)] \mathcal{D}I= \int \delta[\ddot{x}(t)]
\mathcal{D}x= \delta(X\!-v_0T).
\end{equation}
The above relation (arguably the easiest path integral known)
establishes the ``propagator" for the process. Using
Eqs.(\ref{EqnPn}), (\ref{FeynObs}), and two applications of
Eq.(\ref{e-GSpropagator}), the first order term is the sum over
all paths with a single event and is given by:
\begin{eqnarray}\label{P1}
\lefteqn{\nonumber P_1(X,T)= \frac{1}{s\sqrt{2\pi}}\int_0^T dt_1
\int_{-\infty}^{+\infty} dx_1 \int_{-\infty}^{+\infty} dI_1}  \\
& &  \;\;\;\;\;\;\;\;\;\;\times\;\; \delta\left( X
-\left((v_0+I_1)(T-t_1)+x_1 \right)\right)  e^{-I_1^2/{2s^2}}
\delta(x_1-v_0t_1-x_0).
\end{eqnarray}
This expression is easy enough to evaluate using the rule
$\delta(ax)=\delta(x)/|a|$ and we will do so shortly.

\subsection{Feynman Rules}
\begin{figure}
\includegraphics[scale=.5]{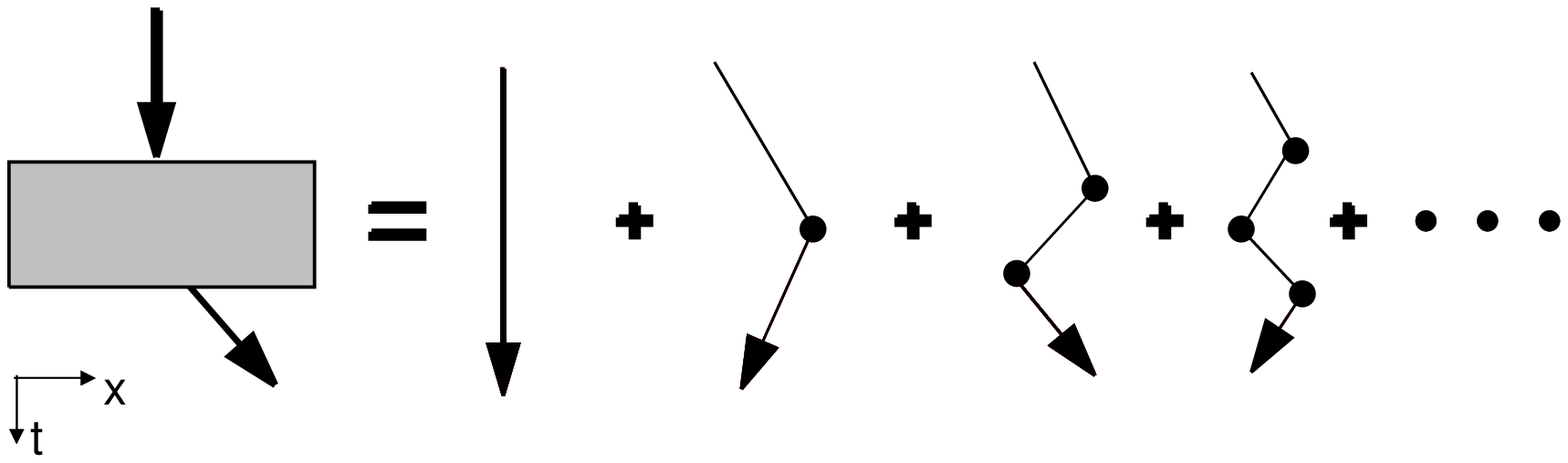}
\caption{\label{GraphSum} The diagrammatic representation of
brownian motion through a medium.}
\end{figure}

Upon writing the sequence of terms $P_1, P_2,...$ we can see the
that the unevaluated integrals (such as Eq.(\ref{P1})) can be
represented by diagrams. The diagrams comprise propagators,
interaction points and the necessary factors to integrate over
intermediate variables. In one-dimension the $n^{th}$ term is
given by:
\begin{equation}\label{GenlEqn}
P_n(X,T)=\int d^2x_n\ldots\int d^2x_1 K_{fn}V_n\ldots
K_{21}V_1K_{1i}
\end{equation}
where $\int
d^2x_i=\int_{-\infty}^{\infty}dx_i\int_0^{t_{i+\!1}}dt_i$. The
specifics of the problem are contained in the propagator $K$ and
the interaction profile $V$. Each term in the above equation, and
the final summation thereof can be represented diagrammatically as
in figure-\ref{GraphSum}.

For the case of a \emph{free }particle undergoing a scattering
process (which conserves momentum), the propagator is:
\begin{equation}\label{Kgs}
K_{i+\!1,i}^{s}=\frac{1}{t_{i+\!1}\!-t_i}\!\!\int_{-\infty}^{+\infty}
\!dI_i
\,\delta\!\!\left(I_i-(\frac{x_{i+\!1}-x_i}{t_{i+\!1}-t_i}-\!\!\sum_{k=1}^i
I_k -v_0)\!\!\right)\!.
\end{equation}
(Note: we will only treat the case of free particles here, so we
will not clutter the super/subscript notation to that effect.) The
interaction points $V^g_i\!=\frac{1}{\sqrt{2\pi
s^2}}e^{I^2_i/2s^2}$ contain the gaussian profile of the $i^{th}$
event.

The propagator can be generalized to other types of motion between
collisions by simply replacing the classical equation of motion as
the argument of the $\delta$-function or other appropriate kernel.

At higher orders, the computation gets increasingly difficult and
a recursion relation between orders of $P_n$ does not exist in
general. This is true in the present case of \emph{gs}-process.
For example to get $P_3^{gs}$ we must repeat the integration over
$x_2$ because $K^{s}_{f3}$ involves $x_2$. Therefore a more
complicated integral over $x_2$ must be performed before the
integration over $x_3$ is done. Despite this complication, the
computation of $P_n^{gs}(X,T)$ can be carried out for arbitrary
$n$, giving:
\begin{equation}\label{Pgs}
P_n^{gs}(X,T)=
 \frac{1}{\sqrt{2\pi
}}\int_0^{\;\;\;T} dt_n\ldots\int_0^{\;\;\;t_2} dt_1
\frac{1}{\xi_n^{gs}}\exp{\left(-\frac{(x-(x_0+v_0T))^2}{2\,{\xi_n^{gs}}^2}\right)}
\end{equation}
where the ``interim variance" ${\xi_n^{gs}}^2$, is given by:
\begin{equation}\label{xigs}
{\xi_n^{gs}}^2=s^2\sum_{i=1}^n(T-t_i)^2.
\end{equation}
As we will see below, the summation over the orders can be easily
carried out in the fourier domain.

Should for some reason, the effect of scattering medium be
artificially stopped after a \emph{fixed} number of events, then
the probability distribution is given by a new class of functions
which we have named: ``Damped exponential integrals". These
functions are described in the appendix-\ref{Dei}.

\subsection{Normalization}
\begin{figure}
\includegraphics[scale=.5]{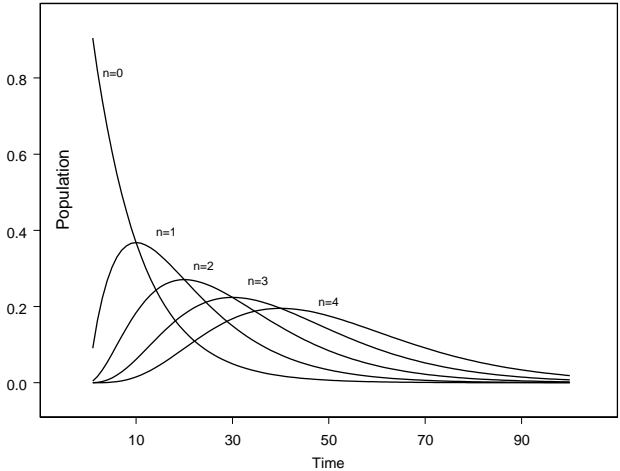}
\caption{\label{RelPops} The relative population Eq.(\ref{Nn}) of
generations $n$, being the number of events they have experienced.
Curves shown here are for $\lambda=0.1$. The peaks occur at
$T=\lambda /n$.}
\end{figure}

We can directly demonstrate the normalization of $P^{gs}(X,T)$ by
integrating Eq.(\ref{Pgs}) over all $X$, and inserting the
resulting terms in Eq.(\ref{Eqn3-3}). Moreover, it is interesting
to note the temporal population of each ``generation" is:
\begin{equation}\label{Nn}
N_n(T)=\frac{e^{-\lambda T}\lambda^n}{n!}\int P_n(X,T) dX=
\frac{(\lambda T)^n}{n!}e^{-\lambda T}.
\end{equation}
Reflecting our initial assumption, this is simply the Poisson
distribution. Summing over all $n$ shows normalization. The
population of generation $n$ will rise and fall according to the
above relation each with a peak at $T=\lambda/n$
(figure-\ref{RelPops}).

Another way to verify the normalization of the gaussian-scatter
solution is as follows. The quantities $N_n(T)$ must satisfy the
simple coupled rate equations:
\begin{eqnarray}
 \dot{N_0}=-\lambda N_0 \\ \nonumber
\dot{N_1}=\lambda N_0-\lambda N_1  \\ \nonumber \ldots \\
\nonumber \dot{N_n}=\lambda N_{n-1}-\lambda N_n.
\end{eqnarray}
It easily verified that the solution for these equations is given
by Eq.(\ref{Nn})

\subsection{Moments and Kurtosis}
We will shortly demonstrate that the exact characteristic function
for the \emph{gs}-process $\Phi_{gs}(k,T)$ can be obtained.
Therefore all moments can be readily obtained by simple
differentiation. However, the ability to obtain the characteristic
function for a given process in closed form is not guaranteed.
Hence we will demonstrate the direct computation of the moments by
summing the moments of each order. We shall specialize to the case
where all initial settings $x_0, v_0, I_0$ are zero. we compute
The $m^{th}$ moment of the $n^{th}$ order distribution and then
sum them according to Eq.(\ref{Eqn3-3}). For the second moment
this is:
\begin{equation}
<\!X^2\!\!>^{gs}_n=
\frac{s^2}{n!}\int_0^T\ldots\int^T_0d^n\tau\sum_{i=1}^n\tau^2_i=\frac{s^2T^{n+2}}{3(n-1)!}
\end{equation}
where $\tau_i=(T-t_i)$. The discrete sum is $n\tau^2$ by symmetry.
The integral on $\tau^2$  is simply $T^3/3$  and the remaining
$n-1$ integrals of measure unity give $T^{n-1}$. Inserting the
result in Eq.(\ref{Eqn3-3}) gives:
\begin{equation}
\nu_{gs}\equiv <\!X^2\!\!>_{gs}= \frac{1}{3}s^2T^2e^{-\lambda
T}\sum_{n=1}^\infty \frac{(\lambda
T)^n}{(n-1)!}=\frac{1}{3}s^2\lambda T^3,
\end{equation}
in agreement with the first order computation in
\cite{FeynPIBook}.

Based on the statement of the problem we can expect
$<\!V^2\!\!>\propto s^2T$. Traditionally the computation of
Brownian motion is within a system in equilibrium with a well
defined temperature. In that case \cite{Chandra,papadop} the
thermal equilibrium constrains the parameter $s$ to the
coefficient of friction and the temperature, and all three to
$<\!V^2\!\!>$. In the present case, no friction exists, therefore
nor a finite temperature as evidenced by the ever increasing
$<\!V^2\!\!>$. The cubic dependence of the variance on observation
time is reminiscent of the diffusion of tracer particles in
turbulent flows. Richardson\cite{richardson} was able to produce
such a behavior for in the context of continuous diffusion only
using a diffusion coefficient which depends on position as
$R^{4/3}$ (alternatively a time dependence of form $T^2$ could
also do it\cite{ShlesingerPhysToday}). More recently, it has been
shown\cite{ShlesingerPRL87} that a Levy-distribution of waiting
times can obviate the need for a space or time dependent diffusion
coefficient in getting several families of supra-linear variances,
one of which includes $T^3$. The above result shows that a Poisson
distribution can also give $T^3$ for the variance.

A similar but more tedious computation for the fourth moment
produces:
\begin{equation}
\chi_{gs}\equiv<\!X^4\!\!>_{gs}= s^4\lambda
T^5(\frac{3}{5}+\frac{\lambda T}{3}).
\end{equation}
The first term is the result of cross-terms like
$(T-t_i)^2(T-t_j)^2$ in the time integrals, whereas the second
term is the result of direct-terms like$(T-t_i)^4$. Ostensibly, it
is the presence of the interference terms that cause deviations
from a normal distribution. Although as expected (from the
central-limit-theorem) the deviation is transient. To quantify
this, we compute the kurtosis for the process:
\begin{equation}
\kappa_{gs}\equiv \frac{\chi_{gs}}{3 {\nu_{gs}}^2}-1=
\frac{9}{5\lambda T}.
\end{equation}
Thus the distribution approaches normality in time. It is,
moreover, worth noting that this result is independent of the
parameter $s$.
\begin{figure}
\includegraphics[scale=.53]{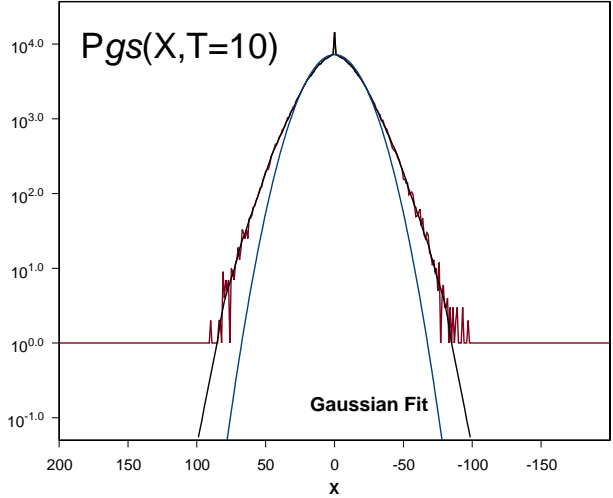}
\caption{\label{P_gsGraph}Demonstration of The Kurtosis(Fat-tails)
as calculated for the \emph{gs}-process at $\lambda T=10$ versus
the computer simulation of the \emph{gs}-process (The noisy line).
The fit of the latter to a normal distribution is also shown. As
expected from the central-limit theorem, the fit to a normal
distribution progresses over time from the peak to the tails.
Unlike the exact solution, the ballistic peak is not present in
the normal distribution.}
\end{figure}

\subsection{The Characteristic Function}
\begin{figure}
\includegraphics[scale=.53]{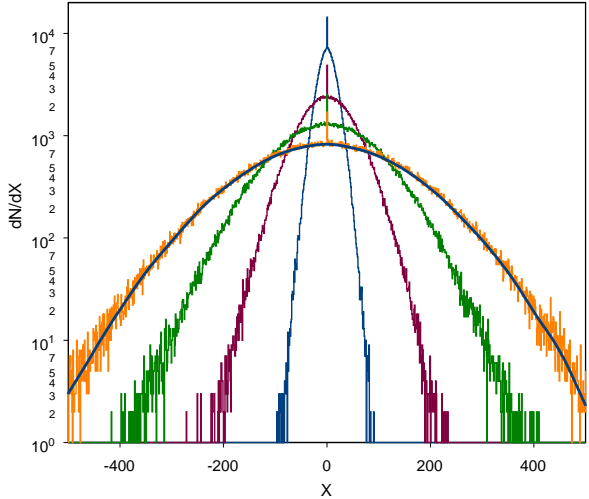}
\caption{\label{P_gsGraphs}$P_{gs}(X,T)$ as calculated by a
numerical simulation of the \emph{gs}-process. The one-dimensional
distribution is shown at different observation times $T$. For
comparison, at the corresponding time, the prediction of the
theory (numeric inverse transform Eq.(\ref{Phi_gs})) is superposed
on the widest distribution. Only the amplitude of the theoretical
curve has been adjusted to match the simulation. The spike at
$X=0$ is the ballistic peak corresponding to the zeroth order
theory (not shown).}
\end{figure}

Having arrived at a solution for $P_n^{gs}(X,T)$ (Eq(\ref{Pgs})),
the characteristic function is easily found by exploiting the
symmetry in $\xi^{gs}$ in the fourier transform of each
$P_n^{gs}$. The symmetry allows us to convert the sequence of
connected integrals into the $n^{th}$ power of a single integral
from 0 to $T$ with an added $1/n!$ pre-factor. We can then sum
them according to Eq.(\ref{Eqn3-3}). The result is:
\begin{equation}\label{Phi_gs}
\Phi_{gs}(k,T)= e^{-\lambda T}\exp{\left( \sqrt{\frac{\pi}{2}}
\frac{\lambda}{ks} \: E\!r\!f
\!\!\left(\frac{ksT}{\sqrt{2}}\right) \right)}.
\end{equation}
Both the second and fourth moments can readily be verified by
repeated differentiation of $\Phi_{gs}(k,T)$ with respect to $k$.
The distribution $P_{gs}(X,T)$ is shown in figures-\ref{P_gsGraph}
and \ref{P_gsGraphs} versus a computer simulation.

If it should happen that the deterministic equation of motion of
the projectile is of the general type, $x(t)\propto v_0t^\alpha
(0<\alpha\leq 1)$, then the result is a simple redefinition of
Eq.(\ref{xigs}) and for certain rational values of $\alpha$ the
characteristic function can be easily computed. The result is that
the variance will be proportional to $T^{1+2\alpha}$ and fourth
moment will be such that the kurtosis remains proportional to
$1/T$, independently of both $\alpha$ and $s$. This is consistent
with the Central Limit Theorem.

\subsection{Higher Dimensions}
For the case of a homogeneous and isotropic medium, and where the
scattering profile has no preferred axis, it is easy to verify
starting from the fundamentals of the theory, that the answer is
found by replacing the scalar wave-number $k$, with the magnitude
of the wave-vector. In case such rotational symmetry exists, by
applying the inverse fourier transform we arrive at the
probability density integrated over a D-dimensional spherical
shell at $R$:
\begin{equation}\label{DdimFT}
P(R)=\frac{R}{(2\pi R)^{D/2} }\int_0^\infty \!\!J(D\!/2-1,kR)\;
\Phi(k) \;k^{D\!/2} dk.
\end{equation}
where, $R=|\mathbf{X}|$, and $J(D\!/2-1,kR)$ is the Bessel
function of the first kind and of order $(D\!/2\!-\!\!1)$. Thus,
if the \emph{gs}-process is applicable, we can insert the $D^{th}$
power of the characteristic function $\Phi_{gs}$ directly into the
above relation.

Alas the \emph{gs}-process is not applicable to photons because
the postulate of relativity is not satisfied by the infinite tail
of the Gaussian interaction kernel and the change in momentum
after each scattering event is not independent of those along
other dimensions. We will treat such constrained interactions
separately.

\section{The Effect Of Noise}\label{s-Noise}
If it should be that instead of simple traversal of the medium the
particle \emph{diffuses} in between collisions (``noisy paths").
We can show that insight provided by the imagery of paths allows
us to see our way through this added effect. Suppose
$P_{gsd}(y,T)$ is the probability density of arriving at $(y,T)$
when diffusion is present. arriving at position $y$ can proceed
along an infinite number of paths which would have culminated at
different $x$'s if diffusion were not present. If we let
$y\!=\!x\!+\!z$ then $z$ is the component of motion due to noise.
Thus the $P_{gsd}(y,T)$ is the sum of all pairs $(x,z)$ which
satisfy the above constraint:
\begin{eqnarray}
\lefteqn{\nonumber P_{gsd}(y,T)= \int \!\!dx\!\int\!\! dz\,
P_{gs}(x,T)P_d(z,T)\delta(y-(x+z))} \\ & &
\;\;\;\;\;\;\;\;\;\;\;\;\;\;=\int\!\! dx \,P_{gs}(x,T)P_d(y-x,T).
 \end{eqnarray}
From here it can be easily shown that the effect can wholly be
incorporated into a new propagator $K^{sd}$. This is what we would
have expected based on the previously alluded path imagery. The
new propagator (itself obtainable by the path integral method
\cite{wiegel}) is simply a broadened form of $K^{s}$
\begin{equation}
\!\!\!\!\!\!\!\! K_{i+1,i}^{sd}=\frac{1}{\sigma
\sqrt{2\pi(t_{i+1}-t_i)}}\int_{-\infty}^{+\infty} dI_i
\exp{\left(-\frac{(I_i-(\frac{x_{i+1}-x_i}{t_{i+1}-t_i}-\sum_{k=1}^i
I_k -v_0))^2}{2\sigma^2/(t_{i+1}-t_i)}\right)}
\end{equation}
where $\sigma^2$ is the diffusion coefficient. The computation of
$P_{gsd}(y,T)$ proceeds as before. Computation will show that in
presence of diffusive movement between scattering events, the
diffusion acts in parallel to the process and thus simply adds a
net term to the variance of the process. Thus, the ``interim
variance" is:
\begin{equation}
{\xi_n^{gsd}}^2={\xi_n^{gs}}^2+ \sigma^2T.
\end{equation}
And we find for the full variance:
\begin{equation}
\nu_{gsd}\equiv<y^2>_{gsd}=\frac{1}{3}s^2\lambda T^3 + \sigma^2T;
\end{equation}
the fourth moment:
\begin{equation}
 \!\chi_{gsd}\;\equiv\;<y^4>_{gsd}\;=\;s^4\lambda
T^5(\frac{3}{5}+\frac{\lambda T}{3}) +
\sigma^2T^2(3\sigma^2+2s^2\lambda^2T^2);
\end{equation}
and the kurtosis:
\begin{equation}
\kappa_{gsd}=\frac{9s^4\lambda T}{5(\lambda
s^2T^2+3\sigma^2)^2}\,\,_{\overrightarrow{T_{large}}}
\,\,\frac{9}{5\lambda T}.
\end{equation}
We observe that over long times, the effect of diffusion (noise)
becomes insignificant. Finally the characteristic function is
found to be:
\begin{equation}
\Phi_{gsd}(k,T)= e^{-(\lambda +\sigma^2k^2/2)T}\exp{\left(
\sqrt{\frac{\pi}{2}} \frac{\lambda}{ks} \: E\!r\!f
\!\!\left(\frac{ksT}{\sqrt{2}}\right) \right)}.
\end{equation}

In this section we have seen that diffusive intra-event movement
can be represented by a change in the functional form of the
propagator. Other effects such as that of an external force field
can be added to the theory by incorporating the equation of motion
in the argument of the propagator. For an absorptive medium, a
damping factor in the propagator will be necessary.

\section{The Gaussian Reset Process}\label{s-GRProcess}
Consider if instead of scattering, we characterize the events as
``resets" in the particle's momentum. It would be as if the
previous momentum is lost at each event and a new one selected
from a gaussian distribution of width $s$. This change implies
only a different propagator:
\begin{equation}\label{Kgr}
K^{r}_{i+1,i}=\int \frac{dI_n}{t_{i+1}-t_i}\,
\delta\!\!\left(I_i-\frac{x_{i+1}-x_i}{t_{i+1}-t_i}\right).
\end{equation}
Utilizing this propagator in the Eq.(\ref{GenlEqn}), and other
rules we find the probability density function $P^{gr}_n(X,T)$ to
be the same form as the Eq.(\ref{Pgs}) if only we use a different
``interim variance" function as given by:
\begin{equation}
{\xi_n^{gr}}^2=s^2\sum_{i=1}^n(t_{i+1}-t_i)^2.
\end{equation}
Following the same steps as in the \emph{gs}-process, we can show
that in presence of diffusive movement between reset events the
diffusion acts in parallel to the process and thus simply adds a
net term to the ``interim variance":
\begin{equation}
{\xi_n^{grd}}^2={\xi_n^{gr}}^2+ \sigma^2T.
\end{equation}

Unfortunately, for this process, ${\xi_n^{gr}}^2$ does not posses
the symmetry which allowed us to compute the
characteristic-function in closed form for the \emph{gs}-process.
Thus we must compute the various moments by summing the moments of
each $P^{gr}_n(X,T)$ according to Eq.(\ref{Eqn3-3}). This task,
while a little tedious for higher moments, is straightforward. The
variance is found to be:
\begin{equation}\label{nu-gr}
\nu_{gr}\equiv<X^2>_{gr}=\frac{2s^2}{\lambda^2}
\left(2(e^{-\lambda T}-1)+\lambda T( e^{-\lambda T}+ 1)\right).
\end{equation}
In very early times the variance increases as $s^2\lambda T^3/3$
which matches $\nu_{gs}$. Later, the momentum non-conservation in
this process manifests its different character resulting in slower
spreading of the distribution. For this process the variance
becomes linear in $T$ in the long time limit ($\lambda T\gg 1$).

\begin{figure}
\includegraphics[scale=.53]{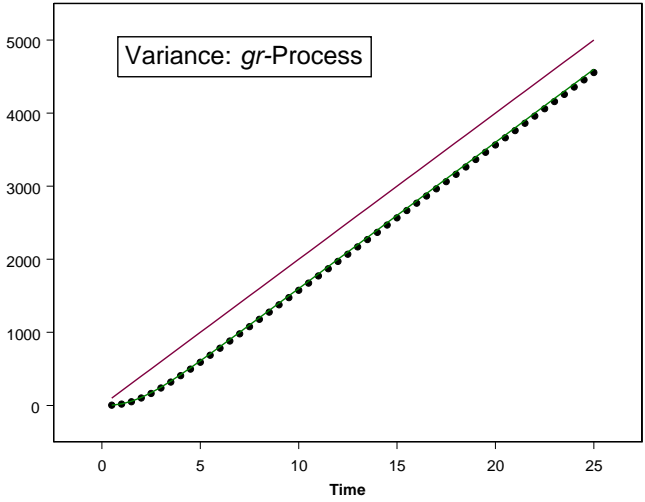}
\caption{\label{nu_gr_Graph}The comparison of the variance as
calculated from a simulation of the \emph{gr}-process (dots),
compared to Eq.(\ref{nu-gr}). The upper solid line is that of
$2s^2T/\lambda$ which is the progression of variance in (Wiener
process) applications such as in Black-Scholes options-valuation.
At early times, the linear approximation grossly overestimates the
\emph{gr}-variance.}
\end{figure}

The fourth moment is:
\begin{eqnarray}\label{chi-gr} \lefteqn{\nonumber \!\!\chi_{gr}\equiv<X^4>_{gr}}\\
& & \;\;\;=\frac{4s^4}{\lambda^4}[e^{-\lambda T}(12+18\lambda T
+9\lambda^2 T^2 +2 \lambda^3T^3)-12\;-6\lambda T+3\lambda^2T^2 ].
\end{eqnarray}

At early times, the fourth moment also behaves like $\chi_{gs}$
increasing proportionally to $T^5$, but later it settles into a
parabolic increase. The kurtosis for the \emph{gr}-process has a
steep fall off at early times, but then it too settles into $\sim
1/\lambda T$ descent:
\begin{equation}
\kappa_{gr}\,\,_{\overrightarrow{T_{large}}} \;2\frac{\lambda
T-4}{(\lambda T-2)^2}.
\end{equation}

Observe that the eventual linearity of the variance is reminiscent
of the Uhlenbeck-Ornstein (\emph{UO}-)process (see
e.g.\cite{Risken},ch.3). The \emph{UO}-process is random flights
(i.e. the Wiener-process) in the presence of continuous friction.
However, continuous friction may not be a plausible mechanism at
the "atomic" scale, where by an "atom" we may mean: an atom of
matter, a node on a network, an individual trader, etc. Hence it
is tempting to to think of the \emph{gr}-process as the
microscopic alternative to the mesoscopic \emph{UO}-process. As a
reminder, both the \emph{UO} and \emph{gr}-processes approach the
limit of the Wiener process, if we turn off friction in the
former($\gamma\rightarrow 0$ but $\gamma\theta$ finite - as
defined below), or allow very large number of collisions per unit
time ($\lambda T \gg 1$) for the latter.

For the \emph{UO}-process, in time, the variance approaches
$\nu_{UO}=2\theta T/m\gamma$, where $\theta$ is the absolute
temperature and $\gamma$ is the coefficient of friction. Thus to
match this with the \emph{gr}-process we must have:
$\theta/m\gamma=s^2/\lambda$. if we arbitrarily set
$\theta/m=s^2$, and $\gamma=\lambda$, then a graph of $\nu_{UO}$
looks quite a bit like that of $\nu_{gr}$ (fig-\ref{nu_gr_Graph}).
However, it is not possible to exactly match the variance of the
\emph{UO}-process to that of the \emph{gr}-process for all times
for any conceivable relation between ($s$,$\lambda$) and
($\theta$,$\gamma$). This stems from the fact that these
parameters represent quite different physical meanings. Still it
is interesting to note that (unlike the \emph{gs}-process) the
\emph{gr}-process has an effective temperature due to the
stabilizing effect of the reset mechanism which acts as as a form
of friction. Thus the choice between the \emph{UO} vs.
\emph{gr}-processes as the appropriate physical model for a given
system lies with the end-user.

We conclude by stating the moral of this section. Even if the
probability distribution (or its characteristic function) may not
be computable in closed form, in many cases, our method allows
access to the exact moments of the distribution.

\section{The Shooting Gallery}
To further demonstrate the versatility of this method we will take
on a problem where a solution by the ``traditional" approach of
stochastic differential equations is difficult. Consider the
carnival game of ``Shooting Gallery". A ``target-duck" starts
moving at the center of a plank with a fixed speed $c$ to the
right. The player is required to shoot the duck. On every hit the
duck reverses direction and continues to retrace its path with
speed $c$. After a time $T$, we need to know the probability
distribution of finding the duck at a given position $X$ within
$dX$. As before, the number of shots is given by the Poisson
distribution of mean $\lambda T$, but otherwise uniformly
distributed over [0,\emph{T}].

The scattering profile (i.e. the distribution from which the next
momentum change is selected) is now comprised of two
$\delta$-functions, each offset by the amount $\pm c$,
respectively. Each lobe of the scattering profile is used in
alternate order. The alternation requirement is an instance of a
situation where the description of the motion does not lend itself
well to analytic description. Because there is a binary
alternation between left-right symmetric profiles we will call
this process the ``symmetric-binary-delta-scattering", or
\emph{sbds}-process.

The propagator for this problem is the same as that of
\emph{gs}-process (Eq.(\ref{Kgs})). Furthermore the result for the
$P_1^{gs}$ when $v_0\neq 0$ (Eq(\ref{P1})) applies if we replace
the interaction kernel (gaussian profile) with $\delta(I_1+2c)$.
Higher orders can easily be written using the Feynman-rules for
this case. Integration over intermediate positions can proceed as
before and the analog to of Eq(\ref{Pgs}) is found to be:
\begin{equation}\label{Pnsbds}
 \!\!\!\!\!P_n^{sbds}(X,T)=
 \int_0^T dt_n\ldots\int_0^{\;\;\;t_2} dt_1 \delta\!\left(X-(v_0-c)T-c \sum_{i=0}^n (-1)^n
(t_{i+1}-t_i)\right)
\end{equation}
where $t_0=0$, and $t_{n+1}=T$. We consider only the initial
condition $v_0=c$ which results in cancellation the second term.
Although higher orders of the above integral require a good bit of
bookkeeping, the actual integration are straightforward if aided
by computer. By extensive use of the rule, $\int_a^b d\tau
\delta(\!\tau-t)f(\tau)=f(t)[\Theta(b\!-t)-\Theta(a\!-t)],$ we
find $P_n^{sbds}(X,T)$ to be:
\begin{eqnarray}
\nonumber \delta (X-c T) \;\;\;\;\;\;\;\;\;\;\;\;\;\;\;\;n=0; \\
\nonumber \frac{(c^2 T^2-X^2)^\frac{n-1}{2} }{2\;c^n\;
(n-1)!!^2}\; B(X,cT) \;\;\;\;n=1,3,5,...; \\\nonumber \frac{(X+
c\; T)(c^2 T^2-X^2)^\frac{n-2}{2} }{2\;c^n\;n!!\;(n-2)!!}\;
B(X,cT)\;\;\;\;
n=2,4,6,...;\\
\end{eqnarray}
where $n!!\!=\!(n/2)!\,2^{n/2}$ ($n$-even, e.g.
$6!!=6\cdot4\cdot2$). The ``light cone" is maintained by the
``box-car" function: $B(X,cT)=\Theta(X+c T)-\Theta(X-c T)$.
Finally, the observed probability density is found by summing
$P_n^{sbds}$ according to Eq.(\ref{Eqn3-3}). To aid the summation
process we can shift the dummy index $n\rightarrow 2n+2$ for even
terms and $n\rightarrow 2n+1$ for the odds. The summed result is:
\begin{equation}\label{Psbds}
\!\!\!\!\!\!\!\!\!\!P_{\!sbds}(X,T)=e^{-\lambda T}\!\left(
\delta(X-c T) +\frac{B(X,cT)}{2c/\lambda}\left[ \frac{(X+c
T)}{\rho c/\lambda}I(1,\rho)+I(0,\rho)\right]\right)
\end{equation}
where $I(k,\rho)$ is the modified Bessel function of the first
kind and of order $k$. $\rho$ is a measure of position within the
light-cone, given by:
\begin{equation}
\rho\equiv\left( \frac{c^2 T^2-X^2}{(c/\lambda)^2} \right)^{1/2}
\end{equation}

\begin{figure}
\includegraphics[scale=.75]{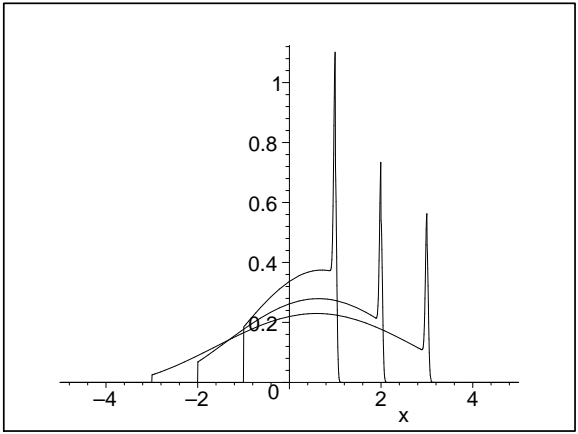}
\caption{\label{FigPsbds2d} The time evolution of the
\emph{sbds}-process for $c=1, \lambda=1$. Here The zeroth order
($\delta(\!X-\!c T)$ in Eq.(\ref{Psbds})) ballistic peaks have
been artificially broadened for plotting.}
\end{figure}

We note that this is in agreement with \cite{Goldstein} derived by
a laborious solution to a coupled set of ``telegrapher's
equations". Figure-\ref{FigPsbds2d} shows samples of $P_{sbds}$
over a relatively early time range for $\lambda=1$ and $c=1$.

The solution above and that in the figure pertain to the initial
condition: $P(\!v_0\!=\!c)\!\!=\!\!1$. In the event
$P(v_0\!=\!-c)\!\neq \!0$ then the solution can easily be
constructed by the superposition
$P(v_0\!=\!c)P_{sbds}(X,T)+P(v_0\!=\!-c)P_{sbds}(-X,T)$.

Finally, we could have modelled the problem with a reset type
propagator as in Eq(\ref{Kgr}). It is easy to show that the
resulting \emph{sbdr}-process will give the same answer as long as
the initial condition $v_0=\pm c$ applies, but not otherwise.

\section{Isotropic Random Flight in $D$-dimensions at an Invariant Speed}
Extending the previous section's model to higher dimensions is of
great practical interest in physical systems. Our method
facilitates the setup without difficulty. We shall denote the unit
vector along the $d^{th}$ component of motion after the $j^{th}$
scattering event as $e_{jd}$. As before, we assume that the
particle moves at a constant speed $c$, and use the propagator of
Eq.(\ref{Kgs}) for a free, momentum-conserving particle. After the
easy integration over intermediate positions we obtain for the
end-to-end propagator of the $d^{th}$ component of the position:
\begin{equation}\label{dDimIint}
\delta\!\left(R_d-(v_{0d}-ce_{0d})T-c \sum_{j\,=0}^N
e_{jd}(t_{j\,+1}-t_j)\right)
\end{equation}
where $t_0=0$, and $t_{N+1}=T$. The above integrand is quite plain
in its statement about the free particle, and could have been
written without the need for integration over many intermediate
coordinates. We can consider two possible initial conditions: 1.
An incident beam where $v_0=c$, or 2. A source emitter where
$v_0=0$. In the former case we can select $\textbf{v}_0=c
\textbf{e}_0$ resulting in cancelling the middle term, but we must
maintain $\textbf{e}_0=(0,0,...,1)$. Alternatively, we can shift
the sum to start from 1 but remember that the position vector is
given by $\textbf{R}=(x,y,...,z-ct_1)$. The latter case is simpler
because we have $\textbf{e}_0=\textbf{0}$ which simply eliminates
the middle term, the sum remains as is, and
$\textbf{R}=(x,y,...,z)$. Either way, the computations do not
depend on the absolute value of the index $j$, hence this
distinction between initial conditions does not come into play
until the very end whence integrating over interaction times.

According to the Feynman rules, we must now add the interaction
kernels and sum over all allowable intermediate directions
$e_{jd}$ as well as times $t_j$. The allowable states for the
particle include only those which maintain a constant speed $c$.
Therefore, for movement in Euclidean space, we must implement the
constraint $\sum_{d=1}^D e_{jd}^2=1$ for all $j$. Thus the full
diagram for the $N^{th}$ order for the D-dimensional scattering
(\emph{Dds}-)process is:
\begin{eqnarray}\label{PnDds}
\lefteqn{ \nonumber \!\!\!\!\!\!\!\!\!\!\!P_{ND}^{Dds}(X,T)=
\prod_{j\,=0}^N\int_0^{t_{j\,+1}}\!dt_j \;\;\;
\prod_{d=1}^D\!\int_{-\infty}^\infty de_{jd}} \\& &
\;\;\;\;\;\;\;\;\;\;\;\times
\;\;\;V(\textbf{e})\;\;\delta\!\left(R_d-c \sum_{j\,=0}^N e_{jd}
(t_{j\,+1}-t_j)\right) \delta\!\left(1-\sum_{d=1}^D e_{jd}^2
\right).
\end{eqnarray}
As a consequence of the nonlinearity of the unit vector
constraints, additional normalization factors will be needed
depending on the explicit dimensionality. Here we will consider
isotropic scattering and thus set $V(\textbf{e})=1$.

In performing these summations we employ a technique which could
also apply to most of the problems we considered previously, but
did not for the sake of illustration of the possibility of direct
integration. The $\delta$-function can be represented as an
unweighted superposition of plane waves. This action, amounting to
a fourier transformation of the integrand, results in the
computation of the characteristic function instead the probability
density. However, in this case the decomposition is especially
necessary. Once decomposed, terms involving a single $e_{jd}$ are
collected and integrated separately. There is, however, one
further complication: An integration over $e_{jd}$ diverges unless
the wave number corresponding to the second $\delta$-function is
always positive. For this reason we choose an alternate (if
obscure) plane wave superposition given by,
$\delta(x)=Re\left(\int_0^\infty e^{i\kappa
x}d\kappa/\pi\right)\!-\!i/{\pi x}$, where the second term is
understood as the principle value. Together with the ordinary
decomposition of the first $\delta$-function, the integrals over
$e_{id}$ involving terms like $i/{\pi x}$ will all vanish. The
remainder of terms with $e_{jd}$ each result in
$\sqrt{\pi/{i\kappa_i}}
\exp{\left(ik_d^2c^2(t_{j\,+1}-t_j)^2/4\kappa_j\right)}$ where the
nested products over $j$ and $d$ apply. In order to carry out the
integration over the $\kappa_j$ we perform the product over $j$.
This conveniently results in the square of the magnitude of the
$D$-vector $\textbf{k}$ in the exponent. Thus we arrive at:
\begin{equation}\label{phi_id}
\phi_D(k,t_j)=\frac{1}{\pi}\int_0^\infty\frac{e^{-\frac{s_j^2+\kappa_j^2}{i\kappa_j}}}{(i\kappa_j/\pi)^{D/2}}
\;\;d\kappa_j
\end{equation}
where, $s_j\equiv kc(t_{j\,+1}-t_j)/2$ and $k=|\textbf{k}|$.
Applying the product over $j$ to $Re(\phi_D)$, and the integration
over all interaction times $t_j$, we arrive at the characteristic
function for the isotropic scattering after $N$-events:
\begin{equation}\label{Phi_Ddis}
\Phi_N^{Ddis}(k)=\prod_{j\,=0}^N \int_0^{t_{j\,+1}}
Re(\phi_D(k,t_j)) dt_j.
\end{equation}
Before proceeding, we note an important property of $\phi_D$:
\begin{equation}\label{phi_id_property}
\phi_{D+2}=-\frac{1}{2s}\frac{\partial \phi_D}{\partial s}.
\end{equation}
That is, the respective characteristic functions for all odd and
separately, all even dimensions are recursively related. For this
reason we need to evaluate the expression in Eq.(\ref{phi_id}) for
$D$=1, and 2 \emph{only}, viz.,
\begin{equation}\label{phi1}
\phi_1(s)=e^{2is};
\end{equation}
and after normalization (i.e. $\phi(0)=1$),
\begin{equation}\label{phi2}
\phi_2(s)=\frac{1}{2i\pi}K(0,2s/i);
\end{equation}
 where $K(0,x)$ is the zeroth-order
modified Bessel function of the second kind.

The probability density is found by performing a $D$-dimensional
inverse fourier transform on $\Phi_N^{Ddis}$. However, because the
$\Phi^{Ddis}_N(k)$ possesses rotational symmetry, the angular
integrals can be performed and the fourier inversion reduces to
Eq.(\ref{DdimFT}).

\subsection{One-dimensional Isotropic Scattering}
\begin{figure}
  \includegraphics[scale=.41]{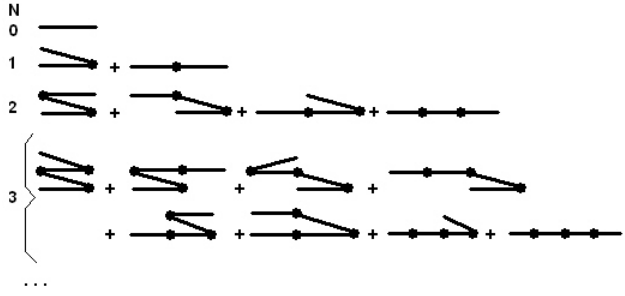}
  \caption{The diagrams that are included in the \emph{1dis}-process.
  The mirror images must also be included, but are not shown}\label{f-1dis-diags}
\end{figure}

Although the motivation for computing the \emph{Ddis}-process was
the generalization of the shooting-gallery problem to higher
dimensions, it turns out that the latter is not the same as the
one-dimensional (\emph{1dis}-)process. This can be most readily
seen via the characteristic function (inserting Eq.(\ref{phi1})
into Eq.(\ref{Phi_Ddis})):
\begin{equation}
 \Phi^{1dis}_N(k,T)=\prod_{j\,=0}^N\int_0^{t_{j\,+1}}\!\!
 cos(kc(t_{j\,+1}-t_j))\;dt_j.
\end{equation}
By setting $N=1$ in the above and fourier transforming the first
order function, we can readily see that the $\delta$-function
found after the integration over the interim coordinates in the
\emph{sbds}-process (Eq.\ref{Pnsbds}) is only one of four we
obtain here. We further note that one of the extra terms
corresponds to a first-order process where no velocity flip takes
place at the event time $t_1$. The remaining two are the parity
($X\!\!\rightarrow -X$) conjugates of the latter two terms. These
features reflect exactly the characteristics which produced
Eq.(\ref{PnDds}). That is we only required that the magnitude of
the speed to remain constant at all times, but allowed all
possible directions. We also allowed isotropic initial conditions
which added the parity conjugate terms. Schematically, the
\emph{1dis}-process encompasses all diagrams in
figure-\ref{f-1dis-diags} (plus their parity conjugates), whereas
at each order, the \emph{sbds}-process includes only the left most
diagram in the figure. In the \emph{1dis}-process, it is as if
after each hit, the duck in the shooting-gallery flips a coin to
decide whether to reverse direction or not.
\begin{figure}
  \includegraphics[scale=.65]{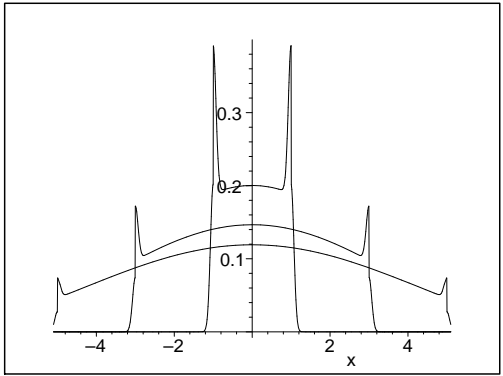}
  \caption{$P_{1dis}$ at $\lambda T=$1, 3, and 5. The ballistic peaks have been artificially broadened for plotting.}\label{f-P1dis}
\end{figure}

While there are $2^N$ diagrams at each order, only (N+1) are
distinct. Other than the ballistic term, present at every order,
we label the remaining $N$ by the index $j$ below. Explicit
computation of the this sub-collection yields:
\begin{equation}\label{PN_1dis} \!\!\!\!\!\!\!\!\!\!\!
P^{1dis}_N(X,T)= \frac{T^N}{2^NN!}
\delta(cT\!-\!X)\!+\!\frac{B(X,cT) N!}{(4c)^N}\sum_{j\,=0}^{N-1}
\frac{(cT\!+\!X)^j (cT\!-\!X)^{N-j-1}}{(n-j)j!^2(N-j-1)!^2}\;
\end{equation}
$B(X,cT)$ is the ``Box-car" function defined previously, which
maintains the light-cone. We can now insert this result in the
summation formula of Eq.(\ref{Eqn3-3}), and add the parity
conjugates, resulting in:
\begin{eqnarray}\label{P_1dis}
\lefteqn{ \nonumber \!\!\!\!\!\!\!\!\!\!\!\!\!\!\!\!
P^{1dis}(X,T)= \frac{1}{2}\{e^{-\lambda T/2}\delta(cT-X)}\\
& & \nonumber \;\;\;\; \;\;+\;\; e^{-\lambda
T}B(X,cT)\sum_{N=0}^\infty\left(\frac{\lambda}{4c}\right)^{N+1}\frac{(cT-X)^N}{N!}
 F\left([-N-1, -N],[1],\rho\right)\}\\
& & \;\;\;\;\;\; + \;\;\frac{1}{2}\{\!X\!\rightarrow\! -X\}.
\end{eqnarray}
Here $\rho\equiv\frac{cT+X}{cT-X}$, and
$F([\alpha,\beta],[\gamma],\rho)$ is the hypergeometric function
of the continuous variable $\rho$. Sample behavior of this
probability density function is shown in figure-\ref{f-P1dis}.

If accuracy is not detrimental, the following simpler expression
for the partial probabilities has the same second moment, and
approximates the fourth moment of $P^{1dis}_N(X,T)$ to within
$5\%$,
\begin{equation}
P^{1dis}_N(X,T)\simeq
\frac{4}{\sqrt{\pi}}\frac{\Gamma\!(\frac{N}{4}\!+\!\frac{3}{2})}{\Gamma
(\frac{N}{4})}\frac{T^{N-1}}{N! (N\!+\!2)}
\left(1-\frac{X^2}{T^2}\right)^{N/4-1}
\end{equation}
applicable for $N\geq 1$. For $T\leq 5$, the full solution
(Eq.(\ref{P_1dis})) without the ballistic term and the light cone,
may also be closely approximated (in units $\lambda=1$, and $c=1$
) with the form: $const/T^2\cdot
\exp\{(T\!+3)(1-\!X^2/(T\!+3)^2)^{1/4}\}$, where $const \sim
.0022$. The approximation can be very good if one is interested
only in small variations in $T$. There, the appropriate value of
the \emph{const} can be found by mere eyeball fitting. For larger
\emph{T}, a gaussian starts to become viable.

\subsection{Two-dimensional Isotropic Scattering}
In two dimensions, the characteristic function is found by
inserting Eq.(\ref{phi2}) into Eq.(\ref{Phi_Ddis}):
\begin{equation}
 \Phi^{2dis}_N(k,T)=\prod_{j\,=0}^N\int_0^{t_{j\,+1}}\!\!
 J(0,kc(t_{j\,+1}-t_j))\;dt_j
\end{equation}
where $J(0,x)$ is the zeroth order Bessel function of the first
kind. The fourier inversion is made possible via Eq.(\ref{DdimFT})
with the kernel $kJ(0,kR)/2\pi$. For the source emitter
configuration, explicit computation yields:
\begin{equation}\label{Pn_2dis}
P_N^{2dis}(R,T)= \frac{\Theta(cT-R) }{2\pi c^N
(N-1)!}\left(c^2T^2-R^2\right)^{(N-2)/2}
\end{equation}
applicable for $N\geq 1$. Summation of orders via the summation
rule Eq.(\ref{Eqn3-3}) yields the full probability density.
\begin{equation}\label{P_2dis}
\! P^{2dis}(X,T)= \frac{e^{-\lambda T}}{2\pi R}\delta(cT-R) +
\frac{e^{-\lambda T}\lambda}{2\pi
c}\frac{\Theta(cT-R)}{\left(c^2T^2-R^2\right)^{1/2}}e^{\frac{\lambda}{c}\sqrt{c^2T^2-R^2}}.
\end{equation}

Note here, that the diffusion limit is found when $cT\!\gg\!R $
rather than the usually assumed limit of $\lambda T\!\gg\! 1$. The
former, however, is in line with the implications of the central
limit theorem, whereas the latter is more of a ``rule of thumb",
which must be used with some care. The above two-dimensional
results agree with those given in \cite{Paasschens}.

\subsection{Three-dimensional Isotropic Scattering}
The 3-dimensional process is likely of greatest practical
interest. Below we specialize to the case of the source emitter.
Using Eq.(\ref{phi_id_property}) we find
$Re(\phi_3)=sin(s_j)/s_j$. Inserting this in Eq.(\ref{Phi_Ddis})
and then into Eq.(\ref{DdimFT}) results in:
\begin{equation}\label{Pn_3dis}
\! P_N^{3dis}(R,T)=\frac{1}{2\pi^2
Rc}\prod_{j\,=0}^N\!\int_0^{t_{j\,+1}}\frac{dt_j}{\tau_j}\int_0^\infty\!\frac{sin(kR)sin(ck\tau_j)}{(ck)^N}\;dk
\end{equation}
where $\tau_j\!\equiv\!(t_{j\,+1}\!-\!t_j)$. As expected, the
zeroth order integral produces the ballistic peak:
$P_0^{3dis}\propto \delta(cT\pm R)$. The first order integrals can
be found by convolution of $\phi_3$ with $P_0^{3dis}$ or, by
direct integration to give:
\begin{equation}
P_1^{3dis}=\frac{\Theta(cT-R)}{4\pi
c^2RT}\,\ln\!\!\left(\frac{cT+R}{cT-R}\right).
\end{equation}

The result of the sine-transform in Eq.(\ref{Pn_3dis}), proves
inaccessible for $N\!>\!1$. However, the moments of the
distribution are easily calculated to arbitrary order. While all
odd moments are zero, the even moments are found by the
application of $(\nabla_k^2)^{m/2}$ to $\Phi_N^{3dis}$, for
$m=0,2,4,...$. For $m=0$ normalization can be verified as
$M_N^{3dis}(0)=T^N/(4\pi N!)$. For $m\geq 2$ The $m^{th}$ moment
is:
\begin{equation}\label{3dis_moments}
M_N^{3dis}(m)=\frac{T^{(N+m)}}{4\,\pi^2 c^m}\frac{2^{m/2}\;(N+1)}{
(N+m)!\;(m-1)!!}\!\sum_{j\,=0}^{m/2-1}\!\!a_j(m)N^j
\end{equation}
where $(m-1)!!\!=\!\!2^{m/2}\Gamma(\frac{m+1}{2})/\sqrt{\pi}$ for
even-$m$ (e.g. $5!!=5\cdot3\cdot1$). Only the set $a_0(m)=
m!^2/(2^m (m/2)!)$ yields to analytic description. The
coefficients $a_j(m)$ for up to the eighth moment are listed in
the table below.

 \vspace{10 pt}
\begin{tabular}{|c|c|c|c|c|}
  \hline
  $a_j(m)$ & \emph{j}=0  & \emph{j}=1  & \emph{j}=2 & \emph{j}=3 \\
  \hline
  \emph{m}=2 & 1 &   &  &  \\
  \emph{m}=4 & 18  & 5  & &  \\
  \emph{m}=6 & 1350  & 1715/3  & 175/3 &  \\
  \emph{m}=8 & 264600  & 137018  &  22785& 1225\\
  \hline
\end{tabular}
\vspace{10 pt}

For $N\!>\!1$, Paasschens\cite{Paasschens} has proposed the
expression:
\begin{equation}
P_N^{3dis}(R,T)\!\simeq
\frac{\Theta(cT\!-\!R)}{\pi^{3/2}c^3}\frac{\Gamma(\frac{3N}{4}\!+\frac{1}{2})}{
\Gamma(\frac{3N}{4})} \frac{T^{N-3}}{N! }
\!\left(\!1\!-\!\frac{R^2}{c^2T^2}\!\right)^{\!\!\!\!\frac{3N}{4}-\!1}
\end{equation}
for the probability density of the 3\emph{dis}-process. This
expression produces the required second moment exactly, the fourth
to within 0.5\%, and the sixth moment is approximated to within
1.5\%. Hence an excellent approximation for many practical
purposes. The total probability density function (via
Eq.(\ref{Eqn3-3})) has also been approximated by
\cite{Paasschens}:

\begin{eqnarray}
\lefteqn{ \nonumber\!\!\!\!\!\!\!\!\! P^{3dis}(R)\simeq
\frac{e^{-\lambda T}}{4\pi R^2}\delta(cT-R)} \\& &
\;\;\;\;\;\;\;+\;\; e^{-\lambda
T}\Theta(cT-R)\frac{(1-R^2/c^2T^2)^{1/8}}{(4\pi
c^2T/3\lambda)^{3/2}}e^{\rho_{3d}}\sqrt{1+2.026/\rho_{3d}}
\end{eqnarray}
where, $\rho_{3d}\equiv \lambda T (1-R^2/c^2T^2)^{3/4}$.

\section{Sample Applications}
The transmission of photons in turbid media is of interest in
medical imaging. During the 1990's increasingly more successful
attempts have been made to model the stochastic movement of the
photons in turbid media
\cite{Durian,Hindus,Perelman95,Poli,Miller,Kaltenbach}. Some of
these have involved forms of path-integration methods while
others, not. However, all of these reports have been limited by
varying forms of approximation, limited dimensionality, and the
like. Some of these approximations pertain to truncated orders of
computation or other more subtle ones such as maintaining the
photons' light-cone\cite{Perelman95} only on average.
Nevertheless, for practical purposes, computations of highly
forward-scattering seem suitable for applications involving
biological tissues. As such, the findings are in reasonably good
agreement within the precision of measurements as reported in
\cite{Perelman95,Winn-Perel98}.

The characteristics of these works have been recounted in a
chronological narrative in \cite{Hindus}. For the case of
isotropic scattering, graphical comparisons of several of these
works to certain exact results can be found in fig-3.3 of
\cite{Paasschens}.

Here we will not consider the mathematically intricate anisotropic
scattering application as it does not make a good illustrative
case. The large body of literature in that realm, nevertheless, is
good evidence of the struggles of usual approximations with early
times ($\lambda T <10$) in stochastic motion. We will however,
consider two simpler popular applications of our results: polymer
chains, and stock option valuation.

\subsection{Flexible Polymer Chains}
A minimal model of a flexible polymer is a chain of links of
constant length $L$ and total length $x$. By ``flexible" we mean
that each bond is free to assume any orientation in space as long
as it remains linked to its two neighbors. Therefore the
probability density distribution of such a chain is a special case
of what we have already considered in the \emph{Ddis}-process
where $c(t_{i+1}-t_i)\!=\!L\!=\!cT/(n+1)$, such that \emph{n}, the
order of computation, is the number of joints. This also means
that the often-difficult time-ordered integration becomes
unnecessary. That is, back in Eq.(\ref{e-Phi_ng}) the probability
of an event at $t_j$ is not uniform over $T$ but restricted to the
single instant $t_j=T/j$. Thus we must use
$\delta(t_j-\frac{jT}{n+1})dt_j$ in place of $\frac{dt_j}{T}$,
leading to the removal of the time integration in expressions such
as Eq.(\ref{Phi_Ddis}) for fixed-sized steps, and resulting in
factors which only affect the normalization. Moreover, because of
the discovery of the recursion rule in Eq.(\ref{phi_id_property})
we need only work out the characteristic function for only one and
two dimensions. All higher dimensions can then be derived from
these, using the recursion relation.

\begin{figure}
\includegraphics[scale=.7]{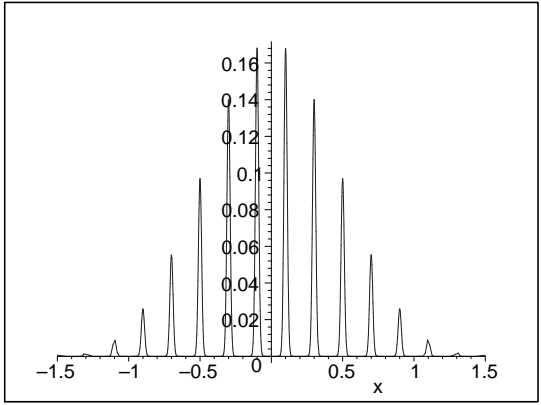}
\caption{\label{Fig-P1dPolymer} The probability density
distribution of Eq.(\ref{e-P1dfc})) for $\ell=21$ links and
$L=0.1$. The peaks have been artificially broadened out for
plotting.}
\end{figure}

The one-dimensional case is likely of limited interest but we
provide the results here for reference. Using Eq.(\ref{phi1}) and
Eq.(\ref{Phi_Ddis})(without the time integration), the
characteristic function for a chain of $\ell$ links is the
$\ell^{th}$ power of $cos(kL)$, and the effective fourier kernel
is $cos(kx)$ for $k\!\in\!(-\infty,\infty)$ and, $x
\!\in\!(-\infty,\infty)$. Writing the cosine as exponentials, the
transform is easily found as a collection of $\delta$-functions
weighted by binomial coefficients:
\begin{eqnarray}\label{e-P1dfc} \lefteqn{\nonumber
\!\!\!\!\!\!\!P^{1df\!c}_\ell(x)=}\\
& &
\frac{1}{2^\ell}\sum_{j=1}^{\frac{\ell+1}{2}}\left(^{\;\;\;\;\;\ell}_{
\frac{\ell+1}{2}-j}\right)(\delta(x\!-\!L(2j\!-\!1))+\delta(x\!+\!L(2j\!-\!1)));
\;\;\;\;
\ell=1,3,... \\
& &\nonumber \frac{1}{2^\ell}\!\left(^{\;\,\ell}_{
\ell/2}\right)\!\delta(x)+\frac{1}{2^\ell}\!\!\sum_{j=1}^{\frac{\ell}{2}}\!\!\left(^{\;\;\,\ell}_{
\frac{\ell}{2}-j}\right)(\delta(x\!-\!2jL)+\delta(x\!+\!2jL));\;\;\;
\ell=2,4,... \;\;.
\end{eqnarray}
It should be apparent that for large $\ell$ the binomial
coefficients will tend toward a Normal distribution enveloping the
discrete spikes (Fig-\ref{Fig-P1dPolymer}). In that limit, the
spikes become indistinguishable and one can approximate
$P^{1df\!c}_\ell(x=mL)$ as a single term given by
$e^{-m^2/2n}/\sqrt{2\pi n}$\cite{Pearson}.

The two-dimensional problem has a famous history\cite{Pearson} but
it is easily addressed using our recursion relation. In
two-dimensions, using Eq.(\ref{phi2} and \ref{Phi_Ddis}) the
characteristic function for a chain of $\ell$ links is the
$\ell^{th}$ power of $J(0,kL)$, and the effective fourier kernel
is $kJ(0,kr)$ for both $k$ and, $r \!\in\![0,\infty)$. Despite the
fact that the same transformation in conjunction with the
time-ordered integration could be worked out in the case of
\emph{2dis} process, we cannot analytically obtain the probability
density function for the end-to-end distance for higher than
second order ($\ell\!>\!3$). The zeroth order contains only a
single link and obviously corresponds to $\delta(L\!-\!r)/2\pi r$,
and the first order is found to be,
$\Theta(2L\!-\!r)/\pi^2r\sqrt{4L^2\!-\!r^2}$. The second order
(3-link) chain has the probability density for the end-to-end
distance, $\textsc{K}(A/\sqrt{L^3r})/\pi^2\sqrt{L^3r}$, for
$r\!\in\![L,3L]$, and, $\textsc{K}(\sqrt{L^3r}/A)/\pi^2A$, for
$r\!\in\![0,L]$. Here, $\textsc{K}$ is the complete elliptic
integral of the first kind, and $A\!=\!(L+r)\sqrt{L+r)(3L-r)}/4$
is the same as the area subtended by the quadrilateral formed by
the the links in the chain and the end-to-end vector of length
$r$. As alluded, higher order distributions remain analytically
inaccessible. But since we have the exact characteristic function,
computing the exact moments at any order is straightforward. Below
we provide up to the eight moment, for chains of up to five links.

\vspace{10 pt}
\begin{tabular}{|c|c|c|c|c|}
  \hline
  $M/L^m$ & \emph{m}=2 & \emph{m}=4 & \emph{m}=6 & \emph{m}=8 \\
  \hline
  \emph{n}=0 & 1   & 1  & 1   & 1 \\
  \emph{n}=1 & 2   & 6  & 20  & 70 \\
  \emph{n}=2 & 3   & 15 & 93  & 639 \\
  \emph{n}=3 & 4   & 28 & 256 & 2716 \\
  \emph{n}=4 &    5& 45 & 545 & 7885 \\
  \hline
\end{tabular}
\vspace{10 pt}

For large-$\ell(\simeq n)$, the distribution has been found by a
number of methods over the last century\cite{Pearson} to approach:
$2r/(L^2n)e^{-r^2/L^2n}$.

The most useful case clearly being that of 3-dimension has been
worked out by Kleinert\cite{kleinert}. We find agreement with this
result as follows. Specializing Eq.(\ref{Pn_3dis}) to the problem
at hand, we find the characteristic function for a chain of $\ell$
links is the $\ell^{th}$ power of $sin(kL)/kL$, and the effective
fourier kernel is $ksin(kr)/2\pi^2r$ for both $k$ and, $r
\!\in\![0,\infty)$. Because the characteristic function is even in
$k$ we can convert the kernel to an exponential and then extend
the lower limit to $-\infty$. If we write $sin^\ell(kL)$ as
combination of exponentials we find an integrand of form
$e^{ikx}/k^{\ell-1}$. The full integral can then performed using
contour integration resulting in\cite{kleinert}:
\begin{eqnarray}\label{e-P3dfc} \lefteqn{\nonumber
P^{3df\!c}_\ell(r)=\frac{1}{2^{\ell+1}(\ell-2)!\pi
L^2r}\sum_{i=0}^{\ell/2}(-1)^i\left(_i^\ell\right)\left(\ell\!-\!2i\!-\!\frac{r}{L}\right)^{\ell-2}
\Theta(L(\ell\!-\!2i)\!-\!r).}\\& &
\end{eqnarray}

Here the large-$\ell(\simeq n)$ limit can be shown to be:
$\frac{12r^2}{nL^3}\sqrt{\frac{3}{\pi n}}e^{-3r^2/2L^2n}$.

\subsection{Financial Options Valuation}
\begin{figure}
\includegraphics[scale=.48]{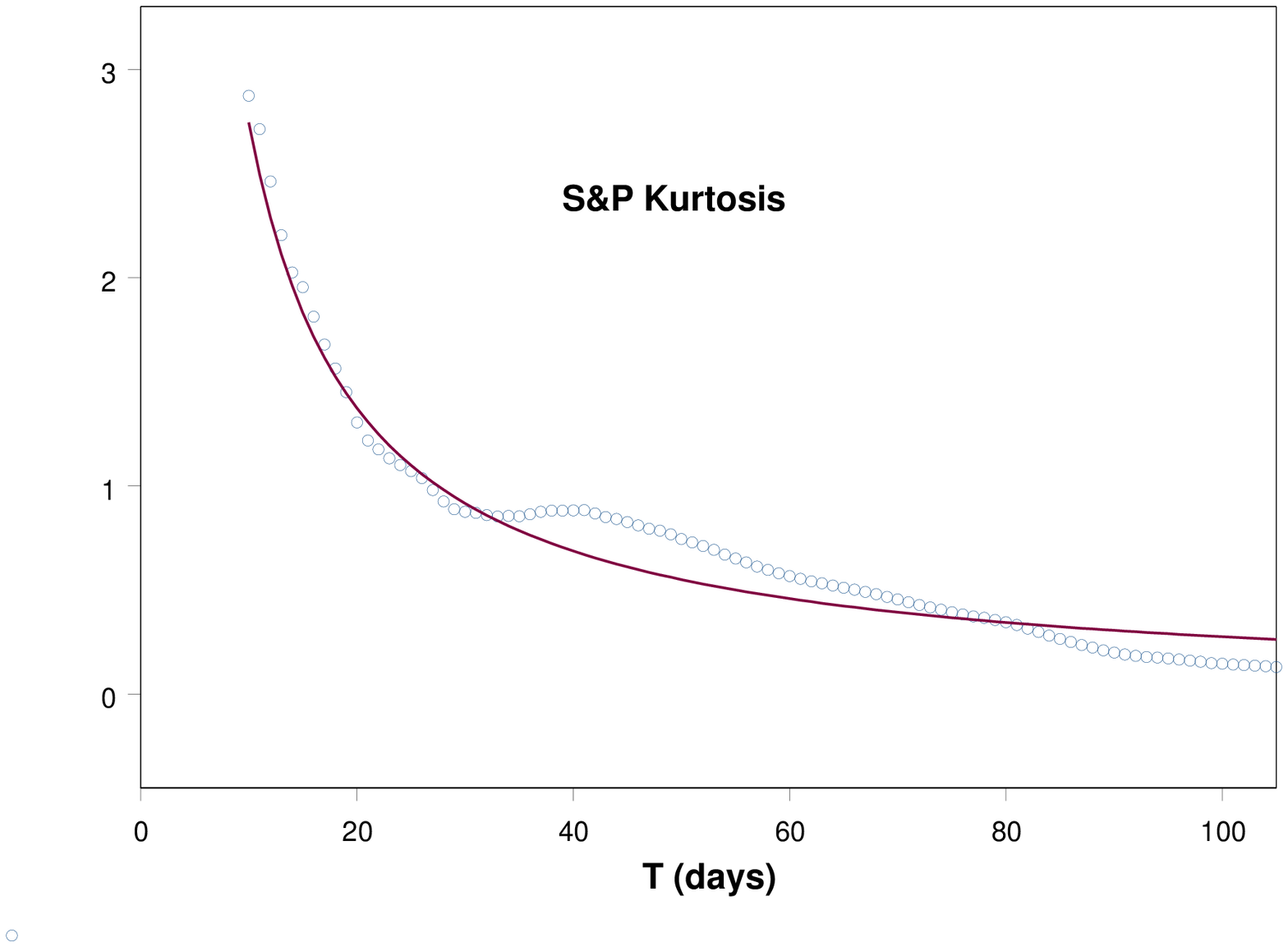}
\caption{\label{S&Pkurtosis} Comparison of the kurtosis for the
daily (Jan-1970 to March-2002) log-returns of the S\&P500 index
vs. a fit of our typical $\kappa\simeq 2/\lambda T$. For
$1/\lambda=13.7 days$ the agreement is good for periods less than
30 days.}
\end{figure}

The problems considered here are especially relevant to the price
movements of financial assets. Economists traditionally assume
that the logarithm of prices have a normal
distribution\cite{Hull}, yet it has always been known this is only
true for very large $\lambda$. While this assumption is fine for
the intended typical gaseous medium where $\lambda \!\sim\!
10^{20}/sec$, the equivalent reset rate for a liquid market is of
order inverse days\cite{NYSECharTime}. Figure-\ref{S&Pkurtosis},
showing the observation-time dependence of the kurtosis over 32
years of S\&P log-returns, clearly indicates a finite $\lambda$
and hence the inappropriateness of the diffusion approximation.

\begin{figure}
\includegraphics[scale=.4]{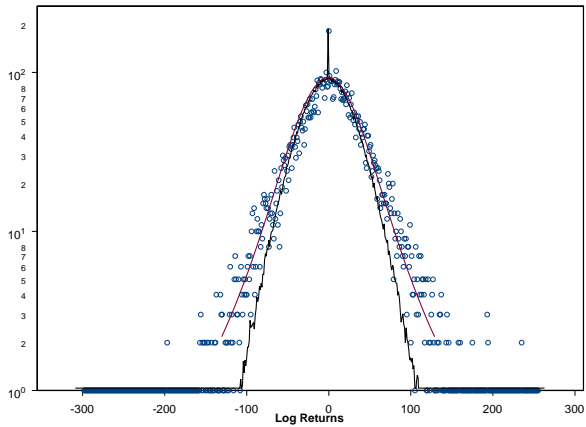}
\caption{\label{f-SPdailyRetsvGRvLevy} The distribution of
log-returns daily (Jan-1970 to March-2002) S\&P500 index vs.
$P_{gr}$, and Levy distributions. While the Levy curve fits the
fat tails much better, a superposition of $P_{gr}$'s can also fit
the fat-tails. Conversely, the Levy-distribution does not have the
ballistic peak; a feature extant in observation though
conveniently ignored by many investigators. Clearly, the correct
physical model which gives rise to the Levy-distribution would
produce the ballistic peak\cite{Zumofen}. The correct relative
amplitude of the ballistic peak to the smooth curve in $P_{gr}$
has not been force-fit.}
\end{figure}

The fact that one cannot fit a single $\lambda$ for different
observation-intervals ($T$) (ranging from minutes to many weeks),
suggests that more than one process govern different time
scales\cite{notLevy}, a notion so common, it is taken for granted
by traders. With more than one $P_{gr}$ curve one can arguably
reproduce the necessary fat-tails as well as Levy-distributions
(figure-\ref{f-SPdailyRetsvGRvLevy}), if only as another
alternative.

In order to see some of the implications of finite $\lambda$, we
present briefly, the valuation of ``options" under the
\emph{gr}-process. Option pricing using path-integrals is quite
commonplace; see\cite{Linetsky,Rosa1,Rosa2,KleinertOpts} to
mention only a very few. The computation of the worth of an
(European-type) call-option for a non-dividend paying underlying
asset was first computed by Black and Scholes \cite{BS}. A
call-option is a contractual right purchased for an agreed upon
premium $C$, which gives the buyer a \emph{right} to purchase a
fixed amount of some asset at a later time $T$ from the seller at
an agreed upon (``strike") price $S$. This is regardless of what
the prevailing market value of the asset might be at $T$.

\begin{figure}
  \includegraphics[scale=.49]{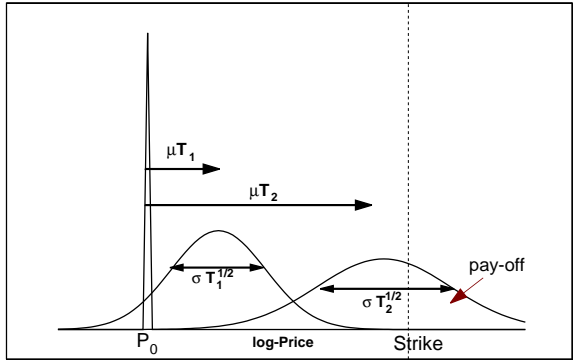}\\
  \caption{The progression of price distribution according to Black and Scholes}
  \label{BSdist}
\end{figure}

Black and Scholes(BS) first assumed that the \emph{logarithms} of
the asset's price follow pure diffusion, with an undetermined
diffusion coefficient $\sigma^2$ (a.k.a., volatility).The
distribution $\delta(ln P-ln P_0)$ reflects the full certainty
that the current price is $P_0$. Thus the distribution will spread
out in time into a Gaussian of variance $\sigma^2T$. They also
assumed the log-price to drift linearly (but no faster) in time
with an undetermined rate $\mu$ (figure-\ref{BSdist}). In this way
they thought they could fix the price probability density
projected for the time $T$ into the future.

The payoff of an option for the buyer would come about if $P_T$
(the market price at  time $T$), is higher than the agreed strike
$S$; the option expires worthless otherwise. If $P_T>S$, the buyer
could exercise the said right and immediately sell the asset on
the open market for a profit of $P_T-S$. Thus a \emph{fair} price
to ask of the buyer is the weighted average of all possible
payoffs:

\begin{equation}\label{BS_Int}
C_{BS} = \frac{e^{-rT}}{\sqrt{2\pi\sigma^2 T}}\int_{\ln S}^\infty
\!\!dx\,\, (e^x-S)e^{-\frac{(x-\ln{P_0-\mu T})^2}{2\sigma^2T}}
\end{equation}
where $r$ is the prevailing interest rate. The reason for the
pre-factor $e^{-rT}$ is that the buyer loses money by missing out
on a steady interest payment since s/he has given up the cash to
buy the call-option. BS then argued: if it is true that no-one can
win consistently at speculating, the drift rate $\mu$ must exactly
be equal\cite{r-sigmaSq} to the guaranteed interest rate $r$. In
this way one of the two arbitrarily introduced parameters was
eliminated. The integral can be written with limits
$x\!\!\in\![-\infty,\infty]$, if we write the integrand with a
step-function. Eq.(\ref{BS_Int}) can then be integrated by parts
resulting in an $Er\!f\!c$ (called $N$ in finance texts). The
final expression is known as the Black-Scholes formula.

Ostensibly, there are assumptions in the BS hypothesis which are
not supported by observation. Conversely, there are many
observations which are not incorporated in the BS model, such as
Levy-like fat tails in the probability
distribution\cite{PhysicaA_v269,mantegna,power3}. The resulting
discrepancies have become deep puzzles in the realm of
finance\cite{FinPhil}. One such puzzle manifested in the pricing
of options is the phenomenon of ``price-skew". In options trading
practice the price of an option cannot really be calculated using
the BS formula, even after eliminating $\mu$, since the
``volatility" ($\sigma^2$) is not known. This parameter may be
inferred from the variance of the detrended price series of the
underlying but using it does not give option prices that match the
observation. Conversely, if the observed option values are
inserted into the BS formula, the implied volatility $\sigma^2$ ,
resulting from solving:
\begin{equation}\label{smile-obs}
C_{obs}=C_{BS}(S,T,P_0,r; \sigma),
\end{equation}
is not the same as that inferred from observation. In fact, using
the observations from different option series (expirations $T$ or
strike prices $S$) one gets different values for $\sigma$, whereas
$C_{BS}$ only allows a unique value for all options. Ostensibly,
there are many more variables which affect $C_{obs}$ in the real
world. The existence of any additional variables or control
parameters immediately implies that for any given $C_{obs}$, the
solution for $\sigma$ is no longer unique. If we take the
\emph{gr}-process as the microscopic basis for the oft-presumed
continuous price diffusion (Wiener) process, it provides just such
a parameter in the form of $\lambda$.

\begin{figure}
\includegraphics[scale=.6]{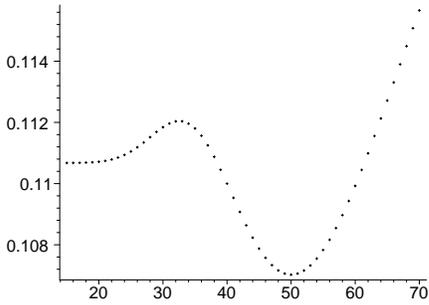}
\caption{\label{skewPlot} The implied volatility (ordinate) vs.
Strike price (abscissa). The two-gausslet ($G_2$) representation
of $P_{gr}$ is used in place of $C_{obs}$ in Eq(\ref{smile-obs}).
Parameter settings are: $\lambda T=10, P_0=50, s=0.1, \mu=0,
r=0.05$. While the features of skew plot remain the same, their
location and span vary to a large extent for different parameters.
For this reason one could get a smile, frown, or smirk relative to
the at-the-money point $S=P_0$. The easiest way to obtain a smirk
or a frown is to allow $\mu\neq 0$, this is consistent with the
findings in \cite{Bouchaud}.}
\end{figure}

It will be illustrative to compute the option skew implied by the
\emph{gr}-process. We will do so only in an approximate way so
that we need not compute the full $P_{gr}$. This will also clearly
demonstrate that the ``fat-tail" of the distribution resulting
from finite values of $\lambda$ is directly linked to the
option-skew phenomenon. We will emulate $P_{gr}$ by matching its
kurtosis to that of a superposition of two gaussians (gausslets):

\begin{equation}
G_2(x;\nu_1,\nu_2)=\frac{e^{-\frac{x^2}{2\nu_1}}}{2\sqrt{2\pi\nu_1}}+
\frac{e^{-\frac{x^2}{2\nu_2}}}{2\sqrt{2\pi\nu_2}}.
\end{equation}
We can now solve for $\{\nu_1, \nu_2\}$ by setting equal the
variance and the fourth moment of $G_2$ to that of $P_{gr}$. For
illustration sake we can take the intermediate time ($\lambda T >
10$) limit of $\nu_{gr}$ Eq. (\ref{nu-gr}), and $\chi_{gr}$ Eq.
(\ref{chi-gr}). In this approximation the solutions are:
\begin{equation}
\nu_{1,2}=\frac{s^2}{\lambda^2}\left(2\lambda T - 4
\mp2\sqrt{(2\lambda T-8)}\right).
\end{equation}
We can see that if we attempt (as traders do in relying on the
BS-model) to fit a single gaussian to the observation, we are
likely to be accepting only one of the above variances. It is
manifest that the implied volatility has a dependence on the time
to expiration $T$ other than the traditional linear term. To
demonstrate the same for the strike price $S$ we use the full form
of $\nu_{gr}$ and $\chi_{gr}$ and solve for the implied gausslet
variances $\{\nu_1, \nu_2\}$. This refined procedure allows us to
get to times as low as $\lambda T \simeq 2$, but to go further we
must match more moments. We produce an option valuation formula as
the half the sum of two Black-Scholes type expressions but using
$\{\nu_1, \nu_2\}$ as the respective variances. Settings this
approximation to $C_{BS}$ and solving for $\sigma$ for different
strikes gives us the the skew curve as shown in the
figure-\ref{skewPlot}.

The application of the method of path-integrals to financial
derivatives' valuations is a very natural approach. Unlike the
case of European (path-independent) options treated above, many
options have values which are path-dependent. Hence the valuation
must take place as part of the path-integration itself
\cite{Linetsky,Rosa1,Rosa2}.

\section{Acknowledgements}
The author gratefully acknowledges M.J.G.Veltman for inspiration
for this work on more levels than methodology; K.Osband for
encouraging the investigation of problems in finance; and
I.M.B.Offen for support and encouragement to complete the work.

\appendix
\section{Damped Exponential Integrals}\label{Dei}
If it should happen that the \emph{gs(d)}-process is terminated
after a \emph{fixed} number of events $n$, then the distribution
(for the truncated-gaussian-scattering with diffusion) is given by
a set of integral functions:
\begin{eqnarray} \lefteqn{\nonumber
\!\!\!\!\!\!\!\!\!P^{tgsd}(x,T)=Dei_n(x,T;\lambda,s,\sigma)}\\
& & \;\;\;\;\;\;\;\;\;\;\;\equiv\frac{1}{N^{Dei}_n} \int_0^T
e^{-\lambda t_n} dt_n\int_0^{t_n}dt_{n-1}\ldots\int_0^{t_2}dt_1
\;\;\frac{1}{\xi^{gsd}_n}
\exp{\!\!\left(-\frac{x^2}{2{\xi^{gsd}_n}^2}\right)}
\end{eqnarray}
where $Dei$ represents the name ``Damped Exponential Integral".
The ``interim variance" is defined as before in Eq.(\ref{xigs}).
The normalization factor is given by:
\begin{equation}
N^{Dei}_n=\frac{\sqrt{2\pi}}{\lambda^n}\left[1-e^{-\lambda
T}\left(\sum_{k=0}^{n-1}\frac{(\lambda T)^k}{k!}\right)\right].
\end{equation}
Thus $Dei_0$ is the familiar normalized Gaussian of width
$\sqrt{\sigma^2 T}$. Further for example, $Dei_1$ is a
superposition of Gaussians of all variances from $\sigma^2T$ to
$(\sigma^2+s^2)T$:
\begin{eqnarray} \lefteqn{\nonumber
\!\!\!\!\!\!\!\!\!Dei_1(x,T;\lambda,s,\sigma)=
\frac{\lambda}{\sqrt{2\pi(1-e^{-\lambda T})}}\int_0^T dt}\\
& & \;\;\;\;\;\;\;\;\;\;\;\;\;\;\times \;\;\frac{e^{-\lambda
t}}{\sqrt{(s^2(T-t)^2+\sigma^2T)}}\;
\exp{\left(-\frac{1}{2}\frac{x^2}{s^2(T-t)^2+\sigma^2T}\right)}.
\end{eqnarray}


\begin{thebibliography}{}
\bibitem{notLevy} We hasten to add that none of the processes
exposed here encompass several of the important observations about
fiancial markets. Most notably the serially correlated absolute
values of returns. These features remain as potential future
applications of the present method.
\bibitem{r-sigmaSq} The actual relationship is
$\mu=r-\sigma^2/2$ since there is a lower return rate if the asset
price is drifting with variance $\sigma^2$
(\emph{op.cit.}\cite{Hull}).
\bibitem{FinPhil} Like other disciplines branching only recently from philosophy, tradiational theories of finance are
founded more on gospelized conjectures, than empirical
observations. As such, many of these ``puzzles" are more semantic
illusions, than genuine mysteries.
\end{thebibliography}
\end{document}